\documentclass{ws-ijmpa2}

\usepackage{amsfonts,amsmath,array,mathrsfs}

\def\be{\begin{equation}}\def\ee{\end{equation}}
\def\cvp{\raise 2pt\hbox{,}}

 \def\tr{\mathop{\rm tr}\nolimits}
 
 \def\Tr{\mathop{\rm Tr}\nolimits}

\def\re{\mathop{\rm Re}\nolimits} 
\def\diag{\mathop{\rm diag}\nolimits}

\def\d{{\rm d}}
\def\nn{{\cal N}} 
\def\llangle{\langle\!\langle}\def\rrangle{\rangle\!\rangle}

 \def\uN{{\rm U}(N)} 
  
\def\La{\Lambda}
\def\mic{\text{mic}}\def\MM{\text{MM}}
 
\def\Rmic{R_{\text{mic}}}
\def\Smic{S_{\text{mic}}}

\def\wmic{W_{\text{mic}}}

\def\a{\boldsymbol{a}}
\def\s{\boldsymbol{s}}\def\g{\boldsymbol{g}}
 \def\m{\boldsymbol{m}}

\def\plb#1#2#3{{\it Phys.\ Lett.\ }{\bf B #1} (#2) #3}
\def\npb#1#2#3{{\it Nucl.\ Phys.\ }{\bf B #1} (#2) #3}

\def\jhep#1#2#3{{\it JHEP\ }{\bf #1} (#2) #3}
\def\prd#1#2#3{{\it Phys.\ Rev.\ }{\bf D #1} (#2) #3}

\def\atmp#1#2#3{{\it Adv.\ Theor.\ Math.\ Phys.\ }{\bf #1} (#2) #3}
\def\cmp#1#2#3{{\it Comm.\ Math.\ Phys.\ }{\bf #1} (#2) #3}

\def\jmp#1#2#3{{\it J.\ Math.\ Phys.\ }{\bf #1} (#2) #3}

\begin{document}

\markboth{Frank Ferrari}
{The Microscopic Approach to Super Yang-Mills}

%
\catchline{}{}{}{}{}
%

\title{THE MICROSCOPIC APPROACH\\ TO $\mathcal N=1$ SUPER YANG-MILLS
THEORIES
}

\author{FRANK FERRARI
}

\address{Service de Physique Th\'eorique et Math\'ematique\\
Universit\'e Libre de Bruxelles and International Solvay Institutes\\
Campus de la Plaine, CP 231, B-1050 Bruxelles, Belgique\\
frank.ferrari@ulb.ac.be}

\maketitle


\begin{abstract}

We give a brief account of the recent progresses in super Yang-Mills
theories based in particular on the application of Nekrasov's
instanton technology to the case of $\nn=1$ supersymmetry. We have
developed a first-principle formalism from which any chiral observable
in the theory can be computed, including in strongly coupled confining
vacua. The correlators are first expressed in terms of some external
variables as sums over colored partitions. The external variables are
then fixed to their physical values by extremizing the microscopic
quantum superpotential. Remarquably, the results can be shown to
coincide with the Dijkgraaf-Vafa matrix model approach, which uses a
totally different mathematical framework. These results clarify many
important properties of $\nn=1$ theories, related in particular to
generalized Konishi anomaly equations and to Veneziano-Yankielowicz
terms in the glueball superpotentials. The proof of the equivalence
between the formalisms based on colored partitions and on matrices is
also a proof of the open/closed string duality in the chiral sector of
the theories.

\keywords{Supersymmetric gauge theories; Gauge-gravity correspondence;
Non-perturbative effects.}
\end{abstract}

\ccode{PACS numbers: 11.30.Pb, 11.15.Tk, 11.25.Tq}

\section{Introduction}	

The study of non-perturbative properties of supersymmetric gauge
theories, by making extensive use of their special properties like
holomorphy, is by now a rather mature field of research, and a great
wealth of results have been obtained over the last 15 years. Many
techniques and ideas have been used (electric-magnetic duality,
integrable systems, mirror symmetry, brane engineering\ldots),
producing an extremely rich and consistent picture of the strongly
coupled regime of super Yang-Mills models. Until recently, the most
general approach was based on an application of the open/closed string
duality to the holomorphic, or chiral, sector of the theory.\cite{GV}
The result of this approach is most elegantly encoded in a planar
matrix model and a suitable glueball superpotential.\cite{DV}

Many attemps have been made to provide gauge theoretic justifications
of the matrix model results. For example, in Ref.~\refcite{ferpot},
the present author provided a proof of the matrix model conjecture (in
the so-called ``one-cut'' case) based on the assumptions of
confinement and the Intriligator-Leigh-Seiberg linearity principle.
These assumptions are of course very natural, but extremely difficult
to justify from first principles. Other works focused on the direct
calculation of the glueball superpotential\cite{Zanon} or the use of
generalized Konishi anomaly equations.\cite{CDSW} However, these
studies were limited to a perturbative analysis, whereas the main
interest in the matrix model is its ability to provide exact
\emph{non}-perturbative results. The conclusion is that in spite of
their great interest in improving our intuitive understanding, all the
above-mentioned approaches fall short in providing valid proofs of the
results.

Our aim in this talk is to present a \emph{first principle,
microscopic} approach to the non-perturbative dynamics of $\nn=1$
gauge theories in the chiral sector, which amounts to computing
directly the relevant path integrals without making any approximation
or assumption.\cite{M1}\cdash\cite{M3} The general observable that we
compute is the expectation value of an arbitrary chiral operator,
\be\label{genvev} \langle\mathscr O\rangle (\g,q)\, .\ee
Such expectation values cannot be corrected in perturbation theory
(which also explains why any perturbative argument that aims at
computing them is doomed to fail), but they do get very non-trivial
and interesting non-perturbative corrections. They depend in general
on the various parameters in the gauge theory action. The gauge
coupling constant enters in the quantum theory through the instanton
factor $q$. There are also various couplings in the tree-level
superpotential and prepotential (see below), that we have denoted
collectively by $\g$ in Eq.~\eqref{genvev}. A typical example of a
chiral correlator is the gluino condensate in the pure $\nn=1$ gauge
theory, which is proportional to a fractional power of $q$,
$\langle\tr\lambda\lambda\rangle = q^{1/N}$. From the knowledge of the
expectation values \eqref{genvev}, one can derive the quantum vacuum
structure and the phases of general $\nn=1$
theories,\cite{fer1}\cdash\cite{phase2} the Seiberg-Witten solution of
the theories with extended supersymmetry,\cite{SW} and actually all
the known exact results in supersymmetric gauge theories.

Our microscopic approach is based on Nekrasov's instanton
technology.\cite{nek1}\cdash\cite{nek5} This was originally developed
for theories with $\nn=2$ supersymmetry, and our main contribution is
to extend it to the $\nn=1$ case (an important early work in this
direction was given in Ref.~\refcite{fucito}). Our results can be seen
as the open string solution of the model, since we start from the
gauge theory action. The solution is expressed in terms of averages
over colored partitions that can be explicitly evaluated. This
description is mathematically very different from the Dijkgraaf-Vafa
recipe\cite{DV} that uses averages over hermitian matrices. Our proof
in Ref.~\refcite{M3}, reviewed below, that the two formalisms yield
exactly the same results for the physical correlators \eqref{genvev}
provides a full justification of the matrix model and equivalently a
proof of the open/closed string duality for the chiral sector of the
$\nn=1$ gauge theories.

An interesting application of our results is to revisit the
perturbatively-derived anomaly equations of Ref.~\refcite{CDSW} and
see if and in what sense they remain true at the non-perturbative
level. As we shall explain, the anomaly equations are non-trivial
dynamical relations from the microscopic point of view (whereas they
correspond to rather trivial identities, the loop equations, in the
matrix model framework). An interesting conceptual result is that,
once these equations have been understood at the non-perturbative
level, the full solution of the model follows.\cite{Cth1,Cth2} The
remaining ambiguities, that are related to the choice of particular
Veneziano-Yankielowicz terms in the glueball
superpotentials,\cite{CDSW} turn out to be completely fixed by general
consistency conditions.

The plan of the talk is as follows. In the next Section, we present
the model on which we focus and explain the main results. In Section
3, we discuss the generalized anomaly equations, emphasizing the
important gap between a perturbative study and the non-perturbative
analysis required to compute the correlators \eqref{genvev}. The power
of the non-perturbative anomaly equations is fully revealed in the
consistency theorem of Ref.~\refcite{Cth1} and \refcite{Cth2} (Theorem
3.2), which we explain. In Section 4, we give more details on the
microscopic formalism. Finally, we discuss possible generalizations
and conclude in Section 5.

\section{The Model and Sketch of the Main Result}

We focus on the paradigmatic example of the $\uN$ theory with one
adjoint chiral superfield $X$. Let us note that there is no difficulty
in considering a more general matter content. If needed, one can also
integrate out the adjoint field by sending its mass to infinity at the
end of the calculations. If we note $W^{\alpha}$ the super field
strength that contains the gauge field and the gluino, the lagrangian
takes the form
\be\label{lag}L=2\re\int\!\d^{2}\theta\,\mathscr W + \text{D-terms},\ee
with
\be\label{Wlag}\mathscr W = -\frac{1}{16\pi^{2}N}\Tr
t''(X)W^{\alpha}W_{\alpha} + N\Tr W(X)\, .\ee
The function $W(X)$ is an arbitrary polynomial tree-level
superpotential, and $t''(X)$ is an arbitrary field-dependent
polynomial gauge coupling. If $W=0$, the theory has $\nn=2$
supersymmetry and $t(X)$ is an arbitrary tree-level prepotential. This
generalized Seiberg-Witten model was studied in Ref.~\refcite{nek5}.
When $W\not = 0$, the theory has only $\nn=1$ supersymmetry. Special
cases are studied in Ref.~\refcite{IM}. It turns out that the r\^oles
of $t$ and $W$ are somehow interchanged in the microscopic and matrix
model formalisms,\cite{M3} which makes the consideration of the
general theory \eqref{Wlag} very natural. The case where $t''=\ln q$
is a constant corresponds to the standard theory.

It is not difficult to show that any correlator \eqref{genvev} can be
written as a sum of products of correlators of the basic variables
\be\label{uvdef} u_{k}=\Tr X^{k}\, ,\quad
v_{k}=-\frac{1}{16\pi^{2}}\Tr W^{\alpha}W_{\alpha}X^{k}\ee
that generate the chiral ring. It is thus convenient to encode the
solution of the theory into two generating functions
\be\label{RSdef} R(z) = \sum_{k\geq 0}\frac{\langle
u_{k}\rangle}{z^{k+1}}\,\cvp\quad S(z) = \sum_{k\geq 0}\frac{\langle
v_{k}\rangle}{z^{k+1}}\,\cdotp\ee

The solution derived in the microscopic formalism takes the following
form. First one introduces averages over the ensemble of colored
partitions (more details are given in Section 4), which physically
label certain field configurations over which the path integrals
localize. If we denote these averages with the symbol
$\lfloor\,\rfloor$, we compute
\be\label{ukmic} u_{k,\,\mic}(\a)=\lim_{\epsilon\rightarrow 0}\lfloor 
\Tr X^{k}\rfloor\, .\ee
The parameter $\epsilon$ corresponds to a deformation of the gauge
theory (the so-called $\Omega$-background) that one must consider in
order to define the measure on colored partitions. The result for the 
original gauge theory are obtained by taking the $\epsilon\rightarrow 
0$ limit. The parameters $\a = (a_{1},\ldots,a_{N})$ correspond to
arbitrary boundary conditions at infinity for the adjoint field $X$,
\be\label{bcX} X_{\infty} = \diag\a\, .\ee
The importance of these boundary conditions will be explained in more 
details in Section 4. It is also possible to express $\lfloor
v_{k}\rfloor$ in terms of the $\lfloor u_{k'}\rfloor,$\cite{fucito,M2}
\be\label{vkmic} v_{k,\,\mic}(\a) =
\frac{N}{(k+1)(k+2)}\lim_{\epsilon\rightarrow
0}\frac{1}{\epsilon^{2}}\bigl(\lfloor\Tr W(X)\Tr X^{k+2}\rfloor -
\lfloor\Tr W(X)\rfloor\lfloor\Tr X^{k+2}\rfloor\bigr)\, .\ee
Generating functions $\Rmic(z;\a)$ and $\Smic(z;\a)$ can then be
defined following Eq.~\eqref{RSdef}. To obtain the physical
correlators, one must fix $\a$ to special values, that are obtained by
extremizing the microscopic superpotential\cite{M1}
\be\label{wmic}\wmic(\a) = \lfloor\Tr W(X)\rfloor\, . \ee
The set of solutions $\a=\a^{*}$ to the equations
\be\label{qem}\d\wmic (\a=\a^{*}) = 0\ee
turns out to be in one-to-one correspondence with the full set of
quantum vacua of the theory.\cite{M1} The physical generating
functions are then given by
\be\label{micphys} R(z) = \Rmic(z;\a^{*})\, ,\quad S(z) =
\Smic(z;\a^{*})\, .\ee

It is interesting to compare the above formalism with the
Dijkgraaf-Vafa matrix model formalism. There, one computes averages,
that we shall denote by $\llangle\,\rrangle$, over hermitian matrices.
The basic identity relates the glueball operators $v_{k}$ to matrix
model averages,
\be\label{vkmac} v_{k,\,\MM}(\s) = N\lim_{\varepsilon\rightarrow
0}\varepsilon\llangle\Tr X^{k}\rrangle\, . \ee
The parameter $\varepsilon$ (not to be confused with $\epsilon$ in the
microscopic formalism) is related to the size of the matrices in the
matrix model, $\varepsilon\sim 1/n$, and thus we take the planar limit
in Eq.~\eqref{vkmac} (although the number of colors $N$ in the gauge
theory is fixed and finite). The arbitrary parameters
$\s=(s_{1},\ldots,s_{r})$ are called the filling fractions. Together
with the integer $r$, $1\leq r\leq\deg W'$, they label the general
solution of the matrix model. The index ``$\MM$'' in \eqref{vkmac}
emphasizes the fact that this is a matrix model average, and it should
not be confused for example with $v_{k,\,\mic}$ defined in
Eq.~\eqref{vkmic}. A formula for $u_{k,\,\MM}(\s)$ can also be given,
which is formally similar to the right hand side of Eq.~\eqref{vkmic}
but with matrix averages replacing averages over colored partitions
and $t''$ replacing $W$ (see Eq.~(2.30) in Ref.~\refcite{M3}). To
$u_{k,\,\MM}$ and $v_{k,\,\MM}$ correspond generating functions
$R_{\MM}(z;\s)$ and $S_{\MM}(z;\s)$ that have been studied extensively
in the literature. Note that \emph{they do not coincide} with the
generating functions $\Rmic(z;\a)$ and $\Smic(z;\a)$ of the
microscopic formalism; they depend on different variables $\a$ and
$\s$, and they have in general different analytic structures (for
example, it is well-known that $R_{\MM}$ and $S_{\MM}$ are two-valued,
algebraic functions of $z$ defined on a hyperelliptic curve; on the
other hand, $\Smic$ is infinitely multi-valued and thus not
algebraic\cite{M3}). A basic conjecture of the matrix model
formalism\cite{DV} is that the physical correlation functions are
obtained for certain values of the filling fractions $\s$, that
correspond, for each value of $r$, to the critical points $\s=\s^{*}$
of a suitable glueball superpotential $W_{\text{glue}}^{(r)}(\s)$,
\be\label{MMeq}\d W_{\text{glue}}^{(r)}(\s=\s^{*}) = 0\, ,\quad 1\leq 
r\leq\deg W'\, .\ee

We can now state our main result. First, the set of solutions of the
$\deg W'$ equations \eqref{MMeq} are in one-to-one correspondence with
the set of solutions of the \emph{single} equation
\eqref{qem}.\cite{M1} Taking into account this correspondence, one can
then show that\cite{M3}
\be\label{MR}\boxed{\Rmic(z;\a^{*}) = R_{\MM}(z;\s^{*})\, ,\quad
\Smic(z;\a^{*}) = S_{\MM}(z;\s^{*})\, .}\ee
These fundamental identities are equivalent to the open/closed string
duality in our case. They imply that when both formalisms are taken
on-shell (i.e.\ when $\a=\a^{*}$ and $\s=\s^{*}$), then the generating
functions computed using colored partitions and matrices coincide.
They also provide the full justification of the matrix model recipe,
since the microscopic formalism is a first-principle approach and the
identification in Eq.~\eqref{micphys} with the physical gauge theory
correlators follows from the basic rules of QFT.

\section{Non-Perturbative Anomalies and Consistency Conditions}

In this Section, we are going to revisit the approach advocated in
Ref.~\refcite{CDSW}, which is based on the study of the gauge theory
equations of motion, but now from a non-perturbative point of view.
For simplicity, we limit the discussion to the standard case for which
$t''$ is a constant.

\subsection{The classical picture}

To understand the nature of the reasoning and of the consistency
conditions we want to use, it is useful to start by analysing the
``trivial'' case of the classical theory. If we note $d=\deg W'$, the 
classical equations of motion take the simple form
\be\label{clem} W'(X) = 0 \propto \prod_{i=1}^{d}(X-x_{i})\, ,\ee
which yields
\be\label{Rcl} R(z) = \sum_{i=1}^{d}\frac{N_{i}}{z-x_{i}}\,\cdotp\ee
The positive integers $N_{i}$ label the classical vacua
$|N_{1},\ldots,N_{d}\rangle$ of the theory and correspond to the
number of eigenvalues of the matrix $X$ that are equal to $x_{i}$. The
gauge symmetry breaking pattern is
$\text{U}(N_{1})\times\cdots\times\text{U}(N_{d})$. An interesting
question to ask is the following: can we write the equations of motion
\eqref{clem} in terms of gauge invariant operators only? This is
essential in view of possible generalizations to the quantum theory. A
complete set of relations on gauge invariant operators derived from
\eqref{clem} is given by
\be\label{clgieq} \Tr \bigl(X^{n+1}W'(X)\bigr) = 0\, ,\quad n\geq -1\,
.\ee
These equations are obtained by considering variations of the form
$\delta X\sim X^{n+1}$, which are generated by the Virasoro-like
operators
\be\label{Vira} L_{n} = -X^{n+1}\frac{\delta}{\delta X}\,\cvp\quad
[L_{n},L_{m}] = (n-m)L_{n+m}\, .\ee
In terms of the generating function $R(z)$, it is straightforward to
check by expanding at large $z$ that the equations \eqref{clgieq} are
equivalent to the requirement that the product
\be\label{prodcl} W'(z)R(z) = N\Delta(z)\ee
must be a polynomial. In other words, the most general solution to
Eq.~\eqref{clgieq} is of the form
\be\label{Rclsolgi} R(z) = \frac{N\Delta(z)}{W'(z)} =
\sum_{i=1}^{d}\frac{c_{i}}{z-x_{i}}\,\cdotp\ee
This is \emph{not} the expected solution, because the $c_{i}$s can be
arbitrary complex numbers, whereas to match the correct solution
\eqref{Rcl} they must be positive integers! So it would seem that the
description in terms of gauge invariant operators (which is also what
one gets in a closed string formalism, where only gauge invariant
objects can be introduced) is missing something. Intuitively, the open
strings (matrix) can be built from the closed strings (gauge invariant
operators) \emph{only when some quantization conditions are
satisfied}.

The fundamental idea to implement these quantization conditions in a
gauge invariant language\cite{Cth1,Cth2} is that since the number of
colors $N$ in the gauge theory is \emph{finite}, then the $u_{k}=\Tr
X^{k}$ are not all independent, but there must exist polynomial
relations of the form
\be\label{relcl} u_{N+p}=P_{p}(u_{1},\ldots,u_{N})\ee
for all $p\geq 1$. The explicit form of the polynomials $P_{p}$ can be
easily found. It turns out that there is a simple but nice algebraic
lemma\cite{Cth2} that states that the relations \eqref{relcl} are
consistent with Eq.~\eqref{Rclsolgi} \emph{if and only if the $c_{i}$ 
are positive integers}.

At the quantum \emph{perturbative} level,\cite{CDSW} one can still
study the consequence of the variations generated by the operator
$L_{n}$ in Eq.~\eqref{Vira} in the path integral. The result
reads\cite{CDSW}
\be\label{anpert1} -N\sum_{k\geq 0}g_{k}u_{n+k+1} +
2\sum_{k_{1}+k_{2}=n}u_{k_{1}}v_{k_{2}} = 0\, ,\ee
where we are using the expansion $W'(z) = \sum_{k\geq 0}g_{k}z^{k}$.
The first term in the left hand side of Eq.~\eqref{anpert1}
corresponds to the classical contribution that we already had in
Eq.~\eqref{clgieq} and the second term is a one-loop anomaly called a
generalized Konishi anomaly. In the quantum theory non-trivial
information also comes from considering the variations generated by
the operators
\be\label{VirJ} J_{n} = \frac{W^{2}}{16\pi^{2}}\frac{\delta}{\delta
X}\, \cvp\quad [L_{n},J_{n}] = (n-m) J_{n+m}\, ,\quad [J_{n},J_{m}] = 
0\, ,\ee
which yield
\be\label{anpert2} -N\sum_{k\geq
0}g_{k}v_{n+k+1}+\sum_{k_{1}+k_{2}=n}v_{k_{1}}v_{k_{2}} = 0\, .\ee

Now comes an important point that was apparently completely overlooked
in the early literature on this subject. The equations \eqref{anpert1}
and \eqref{anpert2} must be supplemented with the constraints
\eqref{relcl}. These constraints are automatically valid to all orders
of perturbation theory. It is then not difficult to show that the only
solutions to Eq.~\eqref{anpert1}, \eqref{anpert2} and \eqref{relcl}
are purely classical, $S(z)=0$ and $R(z)$ given by Eq.~\eqref{Rcl}. In
some sense, we have just rederived, in a very roundabout way using
anomaly equations, the standard perturbative non-renormalization
theorem for chiral operators.

Note that the anomaly equations are very similar to the loop equations
of the matrix model. Actually, Eq.~\eqref{anpert2} precisely coincides
with the loop equations for the $\Tr X^{k}$ in the planar limit, and
this is how the relation \eqref{vkmac} was explained in
Ref.~\refcite{CDSW}. However, we now see that there is a fundamental
difference between the gauge theory and the planar matrix model.
\emph{In the planar matrix model, since the size of the matrix is
infinite, all the variables that enter the loop equations are
independent, and the most general solution is labeled by filling
fractions. In the gauge theory, $N$ is finite and there are
constraints \eqref{relcl}.}

\subsection{The non-perturbative anomaly theorem}

A direct consequence of the above analysis is that the anomaly
equations must be quantum corrected. Otherwise the correlators would
be purely classical! Quantum corrections can be a priori fairly
general, the only obvious constraints coming from global symmetries.
For example, Eq.~\eqref{anpert1} is replaced in the quantum theory by
an equation of the general form
\be\label{anq1} -N\sum_{k\geq 0}g_{k}u_{n+k+1} +
2\!\!\!\!\sum_{k_{1}+k_{2}=n}\!\!\!\!u_{k_{1}}v_{k_{2}}+ \sum_{t\geq
1}q^{t}\Bigl(\sum_{k\geq 0} g_{k}A_{n,k}^{(t)}(u_{p}) + \sum_{k'\geq
0}C_{n,k'}^{(t)}(u_{p})v_{k'}\Bigr) = 0\, .\ee
Similarly, the generators $L_{n}$ and $J_{n}$, Eq.~\eqref{Vira} and
\eqref{VirJ}, and the algebra they generate do get strong quantum
corrections that are computed explicitly in Ref.~\refcite{M2} and
\refcite{M3} and that we describe briefly in Section 4.5.

What is the strongest ``anomaly theorem'' that can be expected to be
valid at the non-perturbative level? The precise statement is as
follows.\cite{Cth1,M3}
\begin{theorem} (Non-perturbative anomaly theorem) It is possible to
absorb the non-perturbative quantum corrections in the anomaly
equations by a suitable redefinition of the variables that enter the
equations.
\end{theorem}
For example, redefinitions of the variables $u_{k}$ of the form
\be\label{redef} u_{k} \rightarrow u_{k} + \sum_{t\geq
1}q^{t}c_{k}^{(t)}(u_{p})\ee
are allowed by the global symmetries if $k\geq 2N$. This simply means
that the operators $u_{k}$ for $k\geq 2N$ are ambiguous in the quantum
theory and a definition in terms of the basic variables must be given.
This is very similar to the ambiguities in defining composite
operators in ordinary QFTs. A precise definition requires the choice
of a particular scheme, and we explain in Section 4.3 that this is
exactly what happens here at the non-perturbative level
(perturbatively, composite chiral operators are unambiguous). One
possible and \emph{perfectly consistent} definition is actually given
by Eq.~\eqref{relcl}, but other choices, corresponding to quantum
corrected relations of the form
\be\label{qrel} u_{N+p}=\mathscr P_{p}(u_{1},\ldots,u_{N},q)\, ,\ee
are also possible. It is crucial to understand that the choice of
relations \eqref{qrel} is \emph{totally arbitrary} and \emph{do not}
correspond to quantum corrections to the chiral ring. Actually, the
ring generated by the $u_{k}$ and $q$ (which is the sector of the
chiral ring of zero R-charge) is the polynomial ring $\mathbb
C[u_{1},\ldots,u_{N},q]$. It is well-known that this ring do not admit
non-trivial deformations that preserve commutativity.

The content of Theorem 3.1 is now clear. For generic definitions of
the variables $u_{k}$, for example with the choice \eqref{relcl}, the
anomaly equations have complicated explicit quantum corrections as in
Eq.~\eqref{anq1}. However, \emph{there exists canonical definitions of
the variables, of the form \eqref{qrel}, that make the quantum
corrections to the anomaly equations implicit.} It is explained in
Section 4.3 that this canonical choice is related to a particular
regularization of the instanton moduli space.

The full proof of Theorem 3.1 is given in Ref.~\refcite{M3} using the
solution from the microscopic formalism. It is however interesting to
make a brief comment on another possible route, that was suggested in
a footnote in Ref.~\refcite{CDSW} and that was tried in the
literature. The idea is to make an ansatz for the possible quantum
corrections to the operators $L_{n}$ and $J_{n}$, for example
\be\label{ansatz} L_{n}\cdot u_{m} = -m u_{n+m} + \sum_{t\geq
1}q^{t}r_{n,m}^{(t)}(u_{p})\, ,\ee
and then try to use Wess-Zumino consistency conditions to constrain
the form of the anomaly equations. However, this approach fails,
because the ansatz \eqref{ansatz} turns out to be wrong. This can be
checked on the exact expressions for the quantum operators $L_{n}$ in
Ref.~\refcite{M3} (see Section 4.5). The problem is that
Eq.~\eqref{ansatz} assumes that the action of the quantum $L_{n}$ on a
chiral operator is given by a polynomial expression. Perhaps
surprisingly, the quantum corrections turn out to be much stronger.

\subsection{The chiral ring consistency theorem}

Let us now explain the main result of Ref.~\refcite{Cth1} and
\refcite{Cth2}. As is well-known, the most general solution to the
anomaly equations depends on a finite number of arbitrary parameters:
the quantum versions of the constants $c_{i}$ that we had at the
classical level in Section 3.1 and also the matrix model filling
fractions. The usual recipe, in the matrix model formalism, is to
postulate that the $c_{i}$ must be positive integers and that the
filling fractions are fixed by extremizing the glueball
superpotentials, Eq.~\eqref{MMeq}. In the microscopic approach, this
is justified by the identities \eqref{MR}, as we develop in Section 4.
However, the philosophy of Ref.~\refcite{Cth1} and \refcite{Cth2} is
to show that the same result can be obtained from Theorem 3.1 by using
algebraic consistency conditions only. 

This is an interesting conceptual result. It shows in particular that
the early point of view on this problem was erroneous. Originally, it
was thought that the anomaly equation part was ``easy'' and that a
full non-perturbative justification, based for example on equations
like \eqref{ansatz}, could be worked out rather straightforwardly. On
the other hand, it was thought that the fixing of the filling
fractions (that correspond to the gluino condensates in the gauge
theory) to their physical values would be extremely difficult to
justify rigorously (the fact that the quantization of the $c_{i}$ was
also highly non-trivial was completely missed in early works). Now the
chiral ring consistency theorem means that the situation is quite the
opposite: once the non-perturbative anomaly theorem is proven (which
turns out to require the full power of the microscopic approach,
invalidating in particular \eqref{ansatz}), the fixing of the filling
fractions (and of the $c_{i}$) follow simply from consistency (and not
from dynamics)!

The proof of the theorem\cite{Cth2} is a generalized version of the
classical analysis given in Section 3.1. The main point is that the
anomaly equations yield an \emph{infinite} set of constraints on a
\emph{finite} set of variables in the gauge theory, because $N$ is
finite (this is very unlike the planar matrix model). Clearly, the
consistency of an infinite number of equations for a finite number of
variables is not obvious at all, and indeed most solutions to the
anomaly equations do not satisfy the requirements. The precise theorem
(derived in Ref.~\refcite{Cth2} in the case of the theory with an
arbitrary number of quark flavors, from which the theory with only the
adjoint field $X$ can be obtained by integrating out the quarks) is as
follows.
\begin{theorem} (Chiral ring consistency theorem) A general solution
to the anomaly equations is not consistent with the existence of
relations of the form \eqref{qrel} between the variables. Consistency
can be achieved only for one particular choice of the polynomials
$\mathscr P_{p}$ in \eqref{qrel} and only when the $c_{i}$s are
positive integers and for special values of the filling fractions.
These special values correspond precisely to the critical points of
the glueball superportentials, as in \eqref{MMeq}, defined to include
uniquely specified Veneziano-Yankielowicz terms.
\end{theorem}
Note that the conditions on $c_{i}$ and the filling fractions are
known to be equivalent to the quantization conditions of the compact
periods of the one-form $R_{\MM}\d z$, which is an elegant and
powerful way to implement the constraints.

Let us finish this Section with a few additional qualitative comments.
The ``philosophy'' of the theorem is that the open strings can be
built from the closed strings only when some quantization conditions
are satisfied in the closed string theory. This is similar to the
quantization of the RR flux in $\text{AdS}_{5}\times\text{S}^{5}$ for
example. In our case, the conditions follow from the extremization of
the flux superpotential (which can be identified with the glueball
superpotential). Another comment is that in the microscopic formalism,
the existence of relations of the form \eqref{qrel} is trivially
implemented, since the operators are constructed from a $N\times N$
matrix $X$. This implies that non-trivial dynamical equations in the
closed string/matrix model framework (which ensures the consistency of
\eqref{qrel} as shown by Theorem 3.2) must be exchanged with trivial
identities in the open string/microscopic formulation. This is exactly
what is found.\cite{M3} On the other hand, off-shell (i.e.~valid for
any $\s$) identities in the closed string/matrix model formulation,
like the generalized anomaly equations, correspond to highly
non-trivial dynamical identities that are only valid on-shell in the
open string/microscopic description. This will be made clear in
Sections 4.4 and 4.5.

\section{The Microscopic Formalism}
\subsection{Nekrasov's technology}

Nekrasov's instanton technology\cite{nek1}\cdash\cite{nek5} is the
crowning achievement of many years of developments in instanton
calculus (see Ref.~\refcite{insta}--\refcite{instc} and references
therein). In a snapshot, this formalism allows to compute any integral
over the moduli space of instantons in the chiral sector, for any
value of the topological charge, and then to sum up exactly the
instanton series (this series always have a finite radius of
convergence). The remarkable mathematical property that underlies this
result is that under certain conditions the integral over the
instanton moduli space localizes over a finite number of field
configurations that are labeled by colored partitions. A generic
formula thus looks like
\be\label{modspaceint}\int\!\d m^{(k)}\mathscr O(\m^{(k)})e^{-\mathcal
S_{\text E}} = q^{k}\sum_{|\vec{\mathsf k}|=k}\mu_{\vec{\mathsf
k}}^{2}\mathscr O_{\vec{\mathsf k}}\, .\ee
The integral in the left hand side in an integral over the moduli
space of instantons of topological charge $k$, $\m^{(k)}$ denoting
collectively the moduli. The expression $\mathscr O(\m^{(k)})$
corresponds to the operator $\mathscr O$, which can be an arbitrary
chiral operator, evaluated on a particular instanton configuration
labeled by $\m^{(k)}$. The right hand side of \eqref{modspaceint} is
proportional to the $k^{\text{th}}$ power of the instanton factor $q$,
and is given by a finite sum over colored partitions $\vec{\mathsf k}$
of size $|\vec{\mathsf k}|=k$. The factor $\mu_{\vec{\mathsf k}}^{2}$
is a measure on the ensemble of colored partitions, and $\mathscr
O_{\vec{\mathsf k}}$ denotes the operator $\mathscr O$ on the
particular instanton configuration labeled by $\vec{\mathsf k}$.

\begin{figure}[pb]
\centerline{\psfig{file=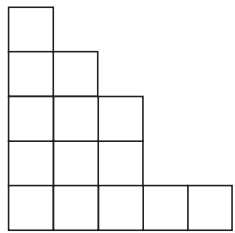,width=3cm}}
\vspace*{8pt}
\caption{The Young tableau associated with the partition
$14=5+3+3+2+1$. \label{f1}}
\end{figure}

In order to get a better understanding of Eq.~\eqref{modspaceint}, let
us give some more details on the ensemble of partitions and of colored
partitions. An ordinary partition $\mathsf k$ of size $|\mathsf k|$ is
simply a decomposition of the positive integer $|\mathsf k|$ into a
sum of positive integers. Thus ordinary partitions are in one-to-one
correspondence with Young tableaux. For example, a partition of the
integer 14 is depicted in Fig.~\ref{f1}. The number of boxes in the
Young tableaux is equal to the topological charge. It is well-known
that Young tableaux are in one-to-one correspondence with irreducible
representations of the symmetric group with $|\mathsf k|$ elements,
and thus it is rather natural to weight a given partition by the
dimension $\dim R_{\mathsf k}$ of this irreducible representation
(this is called the Plancherel measure in the mathematical
literature). Indeed, a careful application of the localization
techniques to the integral in Eq.~\eqref{modspaceint} shows that the
measure factor is given by
\be\label{meas}\mu_{\mathsf k} = \frac{1}{|\mathsf
k|!\epsilon^{|\mathsf k|}}\dim R_{\mathsf k}\, ,\ee
where $\epsilon$ is a deformation parameter that we eventually take to
zero (more is said on this deformation parameter in Section 4.3). The
partitions that enter Eq.~\eqref{modspaceint} are not ordinary
partitions, but colored partitions, and thus the above discussion must
be generalized. A colored partition is simply a collection of $N$
ordinary partitions, $\vec{\mathsf k} = (\mathsf k_{1},\ldots,\mathsf
k_{N})$. The integer $N$ must coincide with the number of colors in
the gauge theory. The measure factor $\mu_{\vec{\mathsf k}}$ is then a
natural generalization of the formula \eqref{meas}. A detailed
discussion can be found for example in the Appendix of
Ref.~\refcite{M2}.

\subsection{Instantons and $\nn=1$ gauge theories}

A calculation based on instantons is non-perturbative, yet it is a
priori valid only at weak coupling. In the case of $\nn=2$
supersymmetry, this is not a limitation. Indeed, $\nn=2$ gauge
theories have a moduli space of vacua. The moduli space always
contains a region where the theory is arbitrarily weakly coupled and
the instanton approximation is exact. The full solution of the theory
can then be obtained by \emph{analytic continuation.} For these
reasons, the instanton technology was originally developed with the
aim of solving $\nn=2$ gauge theories.

The situation for theories with only $\nn=1$ supersymmetry is a priori
much less favorable. Indeed, these theories generically do not have a
moduli space, but a discrete set of vacua. \emph{Some of these vacua
are intrinsically strongly coupled} and thus a direct instanton
analysis is impossible. In particular, observables can be given by
fractional powers of the instanton factor, which is clearly
incompatible with a direct instanton calculation.

The way this problem can be solved was explained in Ref.~\refcite{M1}.
The idea is to make the calculation in two steps. First, one considers
\emph{off-shell}, unphysical, correlators obtained by computing path
integrals with fixed boundary conditions at infinity as in
Eq.~\eqref{bcX}. By choosing appropriately $\a$ (in such a way that
$|a_{i}-a_{j}|\gg\La$, where $\La$ is the dynamically generated scale
of the theory), the path integrals can be forced to be weakly coupled,
and the (unphysical) correlators are then given exactly in terms of
instantons. Since the result must be holomorphic in $\a$, the value of
the correlators for arbitrary $\a$ can be obtained unambiguously by
analytic continuation.

In the second step of the calculation, one computes the exact quantum
superpotential for $\a$. We call this superpotential the microscopic
superpotential $\wmic(\a)$. Using the R-symmetry, $\wmic$ can always
be expressed in terms of the correlators of chiral operators, and can
thus be computed exactly in the instanton approximation using step
one. The superpotential $\wmic$ has a fundamental property that
distinguishes it from any other quantum superpotential previously used
in the literature:\cite{M1} \emph{its critical points are in
one-to-one correspondence with the full set of vacua of the theory}.
In particular, the solutions corresponding to any number of cuts in
the matrix model are obtained as critical points of a single
superpotential, whereas in the matrix model approach one needs a
different glueball superpotential $W_{\text{glue}}^{(r)}$ for each
value of $r$ ($r$ corresponds to the number of cuts), as explained in
Section 2. The fact that all the vacua are found as extrema of a
single superpotential $\wmic$ is of course the signature of the
microscopic nature of our analysis.

We can now understand how the strongly coupled $\nn=1$ vacua are
described with the help of instantons, in an indirect way. We first
compute $\a$-dependent unphysical correlators and $\wmic (\a)$ in the
instanton approximation. The corresponding instanton series have a
finite radius of convergence. We then solve Eq.~\eqref{qem}. Some of
the solutions $\a=\a^{*}$ lie inside the disk of convergence of the
instanton series. These solutions correspond to weakly coupled,
Coulomb-like vacua, for which all the physical correlators can be
expanded as series in $q$ and could actually be computed directly with
instantons without using our two-step procedure. In addition, it turns
out that there are also solutions $\a=\a^{*}$ to Eq.~\eqref{qem} that
lie outside the disk of convergence of the instanton series (they are
typically on the boundary of this disk). \emph{These solutions
correspond to the strongly coupled vacua.} Expanding around such
strongly coupled solutions automatically produce series containing
fractional powers of the instanton factor. For example, this is how
the gluino condensate $\sim q^{1/N}$ is obtained in the pure $\nn=1$
gauge theory.

\subsection{Important technicalities}
\begin{itemlist}
\item Physically, the deformation parameter $\epsilon$ (the so-called
$\Omega$-background) appearing in Eq.~\eqref{meas} provides a nice IR
regulator. The calculation of the scalar correlators $\lfloor\Tr
X^{k}\rfloor$ in the $\epsilon\rightarrow 0$ limit goes essentially as
in the $\nn=2$ theory.\cite{fucito} The glueball correlators
$\lfloor\Tr W^{\alpha}W_{\alpha}X^{k}\rfloor$, which are zero in the
$\nn=2$ case, are much more interesting. Eq.~\eqref{vkmic} shows that
they are related to the next-to-leading order in the small $\epsilon$
expansion of the scalars.\cite{M2,M3} \item Some correlators turn out
to be ambiguous in instanton calculus. For example, if $\Tr X^{k}$ for
$k\geq 2N$ is inserted in the path integral, the result is typically
$0/0$. This singular behavior is due to the small instanton
singularities on the instanton moduli space. The discussion in Section
3.2 gives a clear physical interpretation of these ambiguities: they
correspond to the ambiguities in the definitions \eqref{qrel} of the
variables. Each set of definitions thus corresponds to a choice of
regularization of the instanton moduli space. In Nekrasov's formalism,
we use the non-commutative deformation of the theory to regularize the
moduli space. This particular scheme yields the canonical definition
of Theorem 3.1. 

An interesting extension of these ideas is as follows. The
non-commutative deformation does not work for all gauge groups, but we
conjecture that \emph{there always exists a unique canonical
regularization of the instanton moduli space, for all gauge groups,
corresponding to a choice of variables that make the quantum
corrections in the anomaly equations implicit.}
\end{itemlist}
\subsection{The duality between the colored partitions and the matrix 
model formalisms}

We have developed a new ``open string'' formalism to solve $\nn=1$
theories, based on first-principle path integral calculations. We also
have at our disposal the closed string point of view, based on summing
over hermitian matrices and the Dijkgraaf-Vafa glueball
superpotential. The two formalisms are clearly completely different,
is spite of some formal similarities when one exchange scalar and
glueball operators, colored partitions and matrices, identities and
equations of motion.\cite{M3} 

The fundamental result, which is mathematically summarized by
Eq.~\eqref{MR}, is that when both formalisms are \emph{on-shell}, they
yield the same correlators. Let us give a few more details on how this
equivalence works. 

In the microscopic formalism, we first compute the generating
functions $\Rmic(z;\a)$ and $\Smic(z;\a)$, whereas in the matrix model
formalism one deals with different generating functions
$R_{\MM}(z;\s)$ and $S_{\MM}(z;\s)$. In the matrix model formalism,
extremizing the glueball superpotential is equivalent to the
quantization of the periods of $R_{\MM}\d z$. This means that
\be\label{eq1MM} \d W_{\text{glue}}\sim\oint\! R_{\MM}\d z \mod
2i\pi\mathbb Z\, .\ee
We have explained in Section 3 that the condition $\d
W_{\text{glue}}=0$ ensures the consistency with the open string
formulation of the theory (this is essentially Theorem 3.2). This
means that in the microscopic approach, which is based on the open
string formulation and thus in which the constraints \eqref{qrel} are
trivially satisfied, the conditions derived from $\d
W_{\text{glue}}=0$ should correspond to identities valid off-shell.
This is exactly what is found: one can easily show that
\be\label{eq1mic} \oint\!\Rmic(z;\a)\d z \in 2i\pi\mathbb Z\, ,\ee
\emph{for any} $\a$. Similarly, in the matrix model formalism,
\be\label{eq2MM}\oint\! S'_{\MM}(z;\s)\d z = 0\ee
is a trivial identity valid off-shell (for any $\s$). It follows from
the loop equations of the matrix model. On the other hand, $\Smic'$
does not satisfy an identity like \eqref{eq2MM} for any $\a$, but
rather we have
\be\label{eq2mic} \d\wmic\sim\oint\!\Smic'\d z\, .\ee
Consistency between the formalisms thus comes from the fact that
identities in one formalism, like \eqref{eq1mic} and \eqref{eq2MM},
are exchanged with equations of motion in the other formalism,
\eqref{eq1MM} and \eqref{eq2mic}. Eventually, the full equivalence
when $\a=\a^{*}$ and $\s=\s^{*}$, i.e.~when $\d\wmic=0$ and $\d
W_{\text{glue}}=0$, can be proven.\cite{M3}

\subsection{The anomaly equations}
\label{anosec}

We can now provide a full non-perturbative discussion of the anomaly
equations and of Theorem 3.1. As in any first-principle, microscopic
approach, the anomaly polynomials are expressed as variations of the
microscopic quantum effective action. In the chiral sector we are
discussing, this reduces to variations of the microscopic
superpotential. If we denote by
\begin{align}
\label{An} \mathscr A_{n}(\a) &= -N\sum_{k\geq 0}g_{k}u_{n+k+1,\,\mic} +
2\sum_{k_{1}+k_{2}=n}u_{k_{1},\,\mic}v_{k_{2},\,\mic}\\
\label{Bn} \mathscr B_{n}(\a) &= -N\sum_{k\geq 0}g_{k}v_{n+k+1,\,\mic} +
\sum_{k_{1}+k_{2}=n}v_{k_{1},\,\mic}v_{k_{2},\,\mic}
\end{align}
the anomaly polynomials in the microscopic formalism (compare with
Eq.~\eqref{anpert1} and \eqref{anpert2}), one can construct first
order partial differential operators $\mathscr L_{n}$ and $\mathscr
J_{n}$,\cite{M2,M3} that are the quantum versions of the operators
$L_{n}$ and $J_{n}$ discussed in Section 3, such that
\be\label{anovar} \mathscr L_{n}\cdot\wmic = \mathscr A_{n}\, ,\quad
\mathscr J_{n}\cdot\wmic = \mathscr B_{n}\, .\ee
More precisely, the operators $\mathscr L_{n}$ and $\mathscr J_{n}$
have the form
\be\label{LFform} \mathscr L_{n} =
\sum_{i=1}^{N}\ell_{n,i}(\a)\frac{\partial}{\partial a_{i}}\,\cvp\quad
\mathscr J_{n} =
\sum_{i=1}^{N}j_{n,i}(\a)\frac{\partial}{\partial a_{i}}\, \cvp\ee
where $\ell_{n,i}(\a)$ and $j_{n,i}(\a)$ are given in terms of period
integrals of the one-forms $z^{n+1}\Rmic\d z$ and $z^{n+1}\Smic\d z$
respectively.\cite{M2,M3} 

One can then straightforwardly compute, for example, the quantum
corrections to the perturbative super-Virasoro algebra given by
\eqref{Vira} and \eqref{VirJ}. One then finds in particular that the
quantum algebra does not close on the operators $\mathscr L_{n}$ and
$\mathscr J_{n}$ alone.

We can also compute explicitly the action of $\mathscr L_{n}$ and
$\mathscr J_{n}$ on the variables $u_{m}$ and $v_{m}$. We now
understand why the simple ansatz \eqref{ansatz} cannot work: $\mathscr
L_{n}\cdot u_{m,\,\mic}$ is a perfectly well defined function of $\a$,
but $\a$ itself is a multi-valued function of the $u_{p,\,\mic}$
(because of monodromies in the $\a$-plane, that are strictly similar
to the familiar monodromies in the Seiberg-Witten moduli space of
$\nn=2$ theories). Thus $\mathscr L_{n}\cdot u_{m,\,\mic}$ cannot
possibly be expressed as a polynomial is the $u_{p}$s.

\section{Conclusions}

We have developed a first-principle, text-book like approach to all
the known exact results in $\nn=1$ super Yang-Mills theories. This
provides in particular an explicit and exactly solvable example of the
open/closed string duality. The duality translates mathematically into
a beautiful equivalence between formalisms based on sums over colored
partitions on the one hand and on matrix integrals on the other hand.
It is a unique case where both sides of the duality are now exactly
solved.

Many generalizations are possible. Particularly interesting examples
include turning on some backgrounds (corresponding to higher genus in
the matrix model for instance). Each case must yield a non-trivial and
beautiful equivalence between the open and closed string formulations.

\section*{Acknowledgments}

This work is supported in part by the belgian Fonds de la Recherche
Fondamentale Collective (grant 2.4655.07), the belgian Institut
Interuniversitaire des Sciences Nucl\'eaires (grant 4.4505.86), the
Interuniversity Attraction Poles Programme (Belgian Science Policy)
and by the European Commission FP6 programme MRTN-CT-2004-005104 (in
association with V.\ U.\ Brussels).




\end{document}